\documentclass[prl,twocolumn,twoside,preprintnumbers,superscriptaddress,nofootinbib]{revtex4}
\usepackage{amsmath,slashed}
\usepackage{graphicx,graphics}
\usepackage{dcolumn}
\usepackage[hyperfootnotes=false]{hyperref}
\usepackage{xspace}
\usepackage{mathtools}

\newcommand{\mpi}{M_{\pi}}
\newcommand{\Order}{\mathcal{O}}
\newcommand{\beq}{\begin{equation}}
\newcommand{\eeq}{\end{equation}}
\renewcommand{\Re}{\text{Re}\,}

\newcommand{\diff}{\text{d}}

\newcommand{\F}{\mathcal{F}}
\newcommand{\N}{\mathcal{N}}

\newcommand{\mN}{m_N}

\newcommand{\mc}{m_\chi}

\newcommand{\MeV}{\,\text{MeV}}
\newcommand{\GeV}{\,\ensuremath{\text{GeV}/c^2}}

\newcommand{\bologna}{\affiliation{Department of Physics and Astronomy, University of Bologna and INFN-Bologna, 40126 Bologna, Italy}}

\newcommand{\chicago}{\affiliation{Department of Physics \& Kavli Institute for Cosmological Physics, University of Chicago, Chicago, IL 60637, USA}}

\newcommand{\coimbra}{\affiliation{LIBPhys, Department of Physics, University of Coimbra, 3004-516 Coimbra, Portugal}}

\newcommand{\columbia}{\affiliation{Physics Department, Columbia University, New York, NY 10027, USA}}

\newcommand{\lngs}{\affiliation{INFN-Laboratori Nazionali del Gran Sasso and Gran Sasso Science Institute, 67100 L'Aquila, Italy}}

\newcommand{\mainz}{\affiliation{Institut f\"ur Physik \& Exzellenzcluster PRISMA, Johannes Gutenberg-Universit\"at Mainz, 55099 Mainz, Germany}}

\newcommand{\heidelberg}{\affiliation{Max-Planck-Institut f\"ur Kernphysik, 69117 Heidelberg, Germany}}

\newcommand{\munster}{\affiliation{Institut f\"ur Kernphysik, Westf\"alische Wilhelms-Universit\"at M\"unster, 48149 M\"unster, Germany}}

\newcommand{\nikhef}{\affiliation{Nikhef and the University of Amsterdam, Science Park, 1098XG Amsterdam, Netherlands}}

\newcommand{\nyuad}{\affiliation{New York University Abu Dhabi, Abu Dhabi, United Arab Emirates}}

\newcommand{\purdue}{\affiliation{Department of Physics and Astronomy, Purdue University, West Lafayette, IN 47907, USA}}

\newcommand{\rpi}{\affiliation{Department of Physics, Applied Physics and Astronomy, Rensselaer Polytechnic Institute, Troy, NY 12180, USA}}

\newcommand{\rice}{\affiliation{Department of Physics and Astronomy, Rice University, Houston, TX 77005, USA}}

\newcommand{\stockholm}{\affiliation{Oskar Klein Centre, Department of Physics, Stockholm University, AlbaNova, Stockholm SE-10691, Sweden}}

\newcommand{\subatech}{\affiliation{SUBATECH, IMT Atlantique, CNRS/IN2P3, Universit\'e de Nantes, Nantes 44307, France}}

\newcommand{\torino}{\affiliation{INFN-Torino and Osservatorio Astrofisico di Torino, 10125 Torino, Italy}}

\newcommand{\ucla}{\affiliation{Physics \& Astronomy Department, University of California, Los Angeles, CA 90095, USA}}

\newcommand{\ucsd}{\affiliation{Department of Physics, University of California, San Diego, CA 92093, USA}}

\newcommand{\wis}{\affiliation{Department of Particle Physics and Astrophysics, Weizmann Institute of Science, Rehovot 7610001, Israel}}

\newcommand{\zurich}{\affiliation{Physik-Institut, University of Zurich, 8057  Zurich, Switzerland}}

\newcommand{\paris}{\affiliation{LPNHE, Universit\'{e} Pierre et Marie Curie, Universit\'{e} Paris Diderot, CNRS/IN2P3, Paris 75252, France}}

\newcommand{\freiburg}{\affiliation{Physikalisches Institut, Universit\"at Freiburg, 79104 Freiburg, Germany}}

\newcommand{\lal}{\affiliation{LAL, Universit\'e Paris-Sud, CNRS/IN2P3, Universit\'e Paris-Saclay, F-91405 Orsay, France}}

\newcommand{\naples}{\affiliation{Department of Physics ``Ettore Pancini'', University of Napoli and INFN-Napoli, 80126 Napoli, Italy}} 

\newcommand{\washington}{\affiliation{Institute for Nuclear Theory, University of Washington, Seattle, WA 98195-1550, USA}}
\newcommand{\tudarmstadt}{\affiliation{Institut f\"ur Kernphysik, Technische Universit\"at Darmstadt, 64289 Darmstadt, Germany}}
\newcommand{\helmholtzdarmstadt}{\affiliation{ExtreMe Matter Institute EMMI, GSI Helmholtzzentrum f\"ur Schwerionenforschung GmbH, 64291 Darmstadt, Germany}}
\newcommand{\tokyo}{\affiliation{Center for Nuclear Study, The University of Tokyo, 113-0033 Tokyo, Japan}}

\begin{document}

\title{First results on the scalar WIMP--pion coupling, using the XENON1T experiment}
\author{E.~Aprile}\columbia
\author{J.~Aalbers}\stockholm\nikhef
\author{F.~Agostini}\bologna
\author{M.~Alfonsi}\mainz
\author{L.~Althueser}\munster
\author{F.~D.~Amaro}\coimbra
\author{M.~Anthony}\columbia
\author{V.~C.~Antochi}\stockholm
\author{F.~Arneodo}\nyuad
\author{L.~Baudis}\zurich
% ...
\collaboration{XENON Collaboration}
\email[]{xenon@lngs.infn.it}
\noaffiliation
%non-XENON authors: 
\author{\vspace{-0.3cm}M.\ Hoferichter}\email[]{mhofer@uw.edu}
\washington
\author{P.\ Klos}\tudarmstadt\helmholtzdarmstadt
\author{J.\ Men\'endez}\tokyo
\author{A.\ Schwenk}\tudarmstadt\helmholtzdarmstadt\heidelberg

\begin{abstract}

We present first results on the scalar WIMP--pion coupling from 1 t$\times$yr of exposure with the XENON1T experiment. This interaction is
generated when the WIMP couples to a virtual pion exchanged between
the nucleons in a nucleus. In contrast to most non-relativistic
operators, these pion-exchange currents can be coherently enhanced by
the total number of nucleons, and therefore may dominate in scenarios
where spin-independent WIMP--nucleon interactions are suppressed. Moreover, for
natural values of the couplings, they dominate over the spin-dependent
channel due to their coherence in the nucleus. Using the signal model
of this new WIMP--pion channel, no significant excess is found,
leading to an upper limit
cross section of $6.4\times10^{-46}\ \mathrm{cm}^2$ (90\% confidence level)
at $30\GeV$ WIMP mass.

\end{abstract}

\maketitle

{\it Introduction}.---Profound evidence for the existence of dark matter has been collected
throughout the past 100 years. However, its exact nature remains
elusive~\cite{Bertone:2004pz,Roszkowski:2017nbc}. A large effort is being put into the search for direct detection of
weakly interacting massive particles (WIMPs), which arise as dark
matter particle candidates in various theories.  The search is led
by dual-phase liquid xenon time projection chambers (TPCs) for masses above $~5\GeV$~\cite{Undagoitia:2015gya,Baudis:2016qwx}. The most sensitive experiment, XENON1T, probes spin-independent
(SI) WIMP--nucleon interactions down to $4.1\times10^{-47}\ \mathrm{cm}^2$ for $30\GeV$ WIMP mass~\cite{Aprile:2018dbl}. This limit refers to the SI isoscalar channel, which,
for vanishing momentum transfer $q$, scales
quadratically with the number of nucleons $A$. 
The SI interaction thus yields the
dominant nuclear response, making it the standard search channel
in the field~\cite{Angloher:2015ewa,Armengaud:2016cvl,Akerib:2016vxi,Amole:2017dex,Amaudruz:2017ekt,Cui:2017nnn,Agnese:2017njq,Agnes:2018ves,XMASS:2018bid}.
 
In scenarios where this leading contribution vanishes or is
strongly suppressed, other search channels become important.
Experimentally, this aspect is addressed by dedicated analyses, e.g.,
for spin-dependent (SD) WIMP--nucleon
interactions~\cite{Aprile:2013doa,Uchida:2014cnn,Aprile:2016swn,Fu:2016ega,Akerib:2017kat},
non-relativistic effective field theory (NREFT)
operators~\cite{Schneck:2015eqa,Aprile:2017aas,Xia:2018qgs,Angloher:2018fcs}, or generically
$q$-suppressed responses~\cite{Angloher:2016jsl}.  Contributions
beyond the widely considered SD channel include subleading NREFT
operators~\cite{Fan:2010gt,Fitzpatrick:2012ix,Anand:2013yka}. In addition, a
systematic expansion in the effective theory of QCD, chiral 
EFT~\cite{Epelbaum:2008ga,Machleidt:2011zz,Hammer:2012id,Bacca:2014tla},
valid at the relevant nuclear structure energies and momentum
transfers of the order of the pion mass, reveals a new class of 
contributions referred to as two-body currents. These interactions
proceed by the coupling of the WIMP to a virtual pion exchanged
between nucleons within the nucleus. Such two-body currents that 
occur in the SD channel~\cite{Menendez:2012tm,Klos:2013rwa,Baudis:2013bba}
have already had a significant impact on SD searches, improving substantially the sensitivity
of xenon-based experiments to the SD WIMP--proton cross
section~\cite{Aprile:2013doa,Uchida:2014cnn,Aprile:2016swn,Fu:2016ega,Akerib:2017kat}.

In the SD channel, the inclusion of the leading two-body currents is a
correction to the standard SD response, because it involves the same
WIMP--nucleon coupling. However, in the SI channel the leading two-body
current~\cite{Cirigliano:2012pq,Cirigliano:2013zta,Hoferichter:2015ipa,Hoferichter:2016nvd,Korber:2017ery,Hoferichter:2017olk,Andreoli:2018etf,Hoferichter:2018acd}
cannot be absorbed into a redefinition of the WIMP--nucleon coupling.
Instead, this SI two-body current involves a genuinely new combination of hadronic matrix elements
and Wilson coefficients that describe the interaction of the WIMP with
quarks and gluons~\cite{Goodman:2010ku}. Drawing on the analogy to both SI and SD
WIMP--nucleon interactions, we demonstrate in this paper that these new
couplings can be interpreted as cross sections for a WIMP scattering
off a pion, a channel that has previously not been considered in dark
matter searches.
For natural values of the couplings, this new
WIMP--pion channel dominates over the standard SD channel due to its
coherent nature. 
Here, we present the first results on the scalar WIMP--pion coupling based on the XENON1T experiment.

\begin{figure}[t] 
\centering
\includegraphics[width=\linewidth,clip]{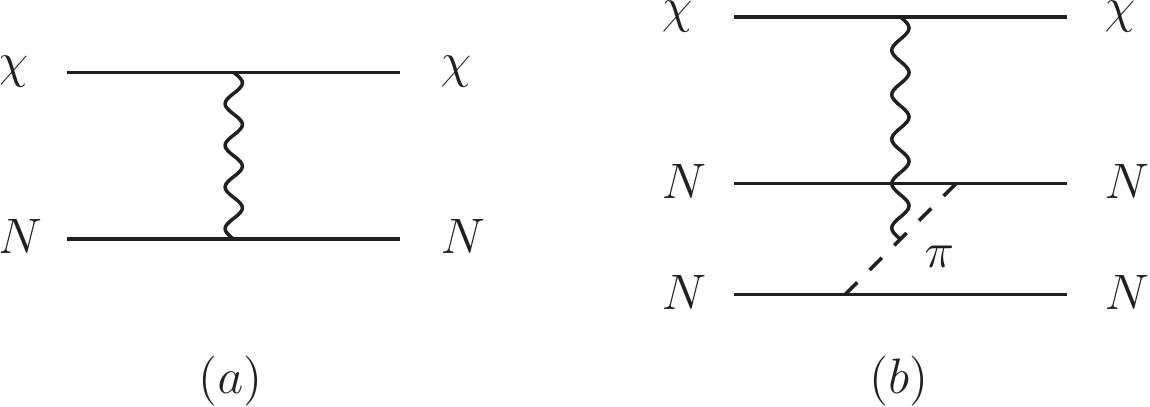}
\caption{Diagrams of WIMP--nucleon interactions. 
Solid lines refer to the WIMP ($\chi$) and nucleon ($N$) fields, wavy
lines to the mediating current, and the dashed line indicates the exchange
of a virtual pion between two nucleons.  $(a)$ Leading WIMP coupling
to one nucleon, $(b)$ two-nucleon contribution from the WIMP coupling
to a pion-exchange current.}
\label{fig:diagrams}
\end{figure}

The key idea is illustrated in Fig.~\ref{fig:diagrams}. Single-nucleon
interactions, both of SI and SD nature, correspond to diagram $(a)$,
where the WIMP $\chi$ interacts only with a single nucleon $N$ within
the nucleus by the exchange of a heavy mediator. Integrating out the
mediator produces effective operators involving the WIMP, quark, and
gluon fields, which together with the hadronic matrix elements
define the single-nucleon cross section that appears as a coefficient
of the WIMP--nucleus rate.  Corrections to this picture emerge from
the fact that a nucleus is a strongly-interacting many-body system,
e.g., mediated by the exchange of virtual pions between two
nucleons. The corresponding coupling of the WIMP through diagram $(b)$
then allows one to interpret limits from the WIMP--nucleus rate as
limits on a WIMP--pion cross section. In the following, we will
consider this mechanism originating from a scalar WIMP--quark coupling
of the form $\bar \chi\chi \bar q q$. For additional details, see Ref.~\cite{Hoferichter:2018acd}.

{\it Theory}.---Analyses of direct detection experiments mostly focus on SI and SD
scattering.
The WIMP--nucleus cross section, $\diff \sigma_{\chi\N}/\diff q^2$, where 
$\N$ indicates the entire nucleus,
depends on the relative velocity of
the WIMP in the lab-frame $v$ and the nuclear spin $J$. 
With nuclear structure factors that encode the response of the nucleus
to the interaction with WIMPs denoted by $\F_\pm^M$~\cite{Fitzpatrick:2012ix} and $S_{ij}$~\cite{Engel:1992bf} for SI and SD scattering, respectively,
this leads to the usual decomposition~\cite{Engel:1992bf,Ressell:1993qm}
\begin{align}
\label{cross_section_dec_simp}
\frac{\diff \sigma_{\chi\N}}{\diff q^2}&=\frac{1}{4\pi v^2}\Big|c_+^M\F_+^M(q^2)+c_-^M\F_-^M(q^2)\Big|^2\notag\\
&+\frac{1}{v^2(2J+1)}\Big(|a_+|^2S_{00}(q^2)+\Re(a_+ a_-^*) S_{01}(q^2)\notag\\
&\qquad+|a_-|^2 S_{11}(q^2)\Big).
\end{align}
Even though the dependence on $q$ itself contains valuable hints for the
nature of the underlying interaction~\cite{Fieguth:2018vob}, the
information about physics beyond the Standard Model (BSM) is fully encoded
in the coefficients $c_\pm^M$ and $a_\pm$. They include both the
coupling of the WIMP to quarks and gluons (Wilson coefficients) and
the hadronic matrix elements that reflect that quarks and gluons are
embedded into nucleons.  The $+$ ($-$) subscript indicates a same-
(opposite-)sign, or isoscalar (isovector), coupling for neutrons and
protons.  In SD scattering it is useful to take $a_+=a_-$, which describes the WIMP coupling to a proton, or $a_+=-a_-$ for the coupling to a neutron.

Most analyses consider the following scenarios.  First, they assume purely
isoscalar SI interactions ($c_-^M=a_\pm=0$), with the WIMP--nucleus
cross section expressed in terms of the SI cross section off a single
nucleon $\sigma_{\chi N}^\text{SI}$:
\beq
\label{eq:SInucleon-coupling}
\frac{\diff \sigma_{\chi\N}}{\diff q^2}=\frac{\sigma_{\chi N}^\text{SI}}{4\mu_N^2v^2}\big|\F_+^M(q^2)\big|^2,\qquad
\sigma_{\chi N}^\text{SI}=\frac{\mu_N^2}{\pi}\big|c_+^M\big|^2,
\eeq
where $\mu_N$ is the WIMP--nucleon reduced mass.
The nuclear structure factor $\F_+^M$ is often
approximated by a Helm form factor~\cite{Helm:1956zz}, but more
sophisticated nuclear calculations are
available~\cite{Vietze:2014vsa}.  Second, one takes a purely SD coupling
($c_\pm^M=0$) with $a_+=a_-$ or $a_+=-a_-$ written in terms of the SD
cross section off a single proton or neutron $\sigma_{\chi N}^\text{SD}$:
\beq
\label{eq:SDnucleon-coupling}
\frac{\diff \sigma_{\chi\N}}{\diff q^2}=\frac{\sigma_{\chi N}^\text{SD}}{3\mu_N^2v^2}\frac{\pi}{2J+1}S_N(q^2),\qquad
\sigma_{\chi N}^\text{SD}=\frac{3\mu_N^2}{\pi}\big|a_+\big|^2,
\eeq
where single nucleons are denoted by $N=\{p,n\}$ and
$S_{p/n}(q^2)=S_{00}(q^2)\pm S_{01}(q^2)+S_{11}(q^2)$.
Out of
these scenarios, the SI response is dominant because all $A$ nucleons
contribute coherently: $\F_+^M(0)^2=A^2$, with $A \sim 130$ for xenon.
In contrast, in the SD channel the response
does not scale with $A$:
$[4\pi/(2J+1)]S_N(0)\sim [4(J+1)/J]\langle {\bf
  S}_N\rangle^2=\Order(1)$ (for nuclei with unpaired nucleons), with
$\langle {\bf S}_{p/n}\rangle$ proton/neutron spin-expectation values
of the nuclear target.\footnote{This estimate does not include contributions from two-body currents to SD scattering, 
which are quantitatively significant especially for the paired species~\cite{Menendez:2012tm,Klos:2013rwa}, but they do not enter coherently.}  
Therefore, SD limits become most relevant if
the SI interactions are either absent or strongly suppressed~\cite{Freytsis:2010ne}. 
In practice, the consideration of limits on $\sigma_{\chi N}^\text{SI}$,
$\sigma_{\chi p}^\text{SD}$, and $\sigma_{\chi n}^\text{SD}$
corresponds to a set of slices through the BSM parameter space, which
is not a complete or unique choice.  For instance, one could also
consider proton- or neutron-only SI cross sections ($c_+^M=\pm c_-^M$,
$a_\pm=0$), which are related to isospin-violating dark
matter~\cite{Kurylov:2003ra,Giuliani:2005my,Chang:2010yk,Feng:2011vu,Cirigliano:2013zta}.

In this paper, we consider the leading contribution beyond SI and SD scattering given in
Eqs.~\eqref{eq:SInucleon-coupling} and \eqref{eq:SDnucleon-coupling}.
For that purpose we use chiral EFT~\cite{Hoferichter:2015ipa}, 
which allows one to derive a more complete set of possible WIMP
interactions with nuclei. When the relevant
momentum transfers are of the order of the pion mass $q\lesssim
M_\pi$, such as in direct detection experiments, chiral EFT predicts
that pions, in addition to nucleons, emerge as relevant degrees of
freedom.  In fact, in chiral EFT nuclear forces are mediated by pion exchanges, and
also the interactions of nuclei with external probes can occur via the
coupling to a pion exchanged between two nucleons.  Such pion-exchange
currents are very well established in electromagnetic and weak
interactions in nuclei (see, e.g.,
Refs.~\cite{Gazit:2008ma,Bacca:2014tla}).

A chiral EFT study of WIMP interactions with nucleons indicates that
pion-exchange currents [see Fig.~\ref{fig:diagrams} $(b)$] enter at
the same order in the chiral EFT power counting as momentum-suppressed
single-nucleon currents~\cite{Hoferichter:2015ipa}. The importance of
pion-exchange currents has been stressed for SD
scattering~\cite{Menendez:2012tm,Klos:2013rwa}, where they lift the strict separation
between proton-/neutron-only couplings. By probing the neutrons
even for $a_+=a_-$ they dramatically increase the sensitivity to
$\sigma_{\chi p}^\text{SD}$ for an experimental target, such as xenon, with
an even number of (mainly paired)
protons~\cite{Aprile:2013doa,Uchida:2014cnn,Fu:2016ega,Akerib:2017kat,Aprile:2016swn}.
Similarly, pion-exchange currents constitute the most important
coherent correction~\cite{Hoferichter:2016nvd,Hoferichter:2018acd}.  Therefore, a minimal
extension of Eq.~\eqref{cross_section_dec_simp} adds a term
corresponding to the WIMP--pion coupling, with a new combination of
Wilson coefficients and hadronic matrix elements, $c_\pi$, together
with a novel nuclear structure factor $\F_\pi(q^2)$:
\beq
\label{cross_section_dec}
\frac{\diff \sigma_{\chi\N}}{\diff q^2}=\frac{1}{4\pi v^2}\Big|c_+^M\F_+^M(q^2)+c_-^M\F_-^M(q^2)+c_\pi \F_\pi(q^2)\Big|^2,
\eeq
without changing the SD interactions. The decomposition in
Eq.~\eqref{cross_section_dec} suggests to consider, in addition to
standard SI/SD analyses, the scenario where $c_\pm^M=a_\pm=0$, leading
to
\beq
\label{eq:pioncoupling}
\frac{\diff \sigma_{\chi\N}}{\diff q^2}=\frac{\sigma_{\chi \pi}^\text{scalar}}{\mu_\pi^2v^2}\big|\F_\pi(q^2)\big|^2,\qquad \sigma_{\chi\pi}^\text{scalar}=\frac{\mu_\pi^2}{4\pi}\big|c_\pi\big|^2,
\eeq
with scalar WIMP--pion cross section $\sigma_{\chi\pi}^\text{scalar}$
and WIMP--pion reduced mass $\mu_\pi$. In analogy to SI/SD limits, the
structure factor $\F_\pi$ then allows one to derive limits for
$\sigma_{\chi \pi}^\text{scalar}$ as a function of the WIMP mass $\mc$.
The corresponding exclusion plot represents another slice in the BSM
parameter space. It becomes relevant for regions where cancellations occur in the leading SI
coupling to nucleons, e.g., in
heavy-WIMP EFT~\cite{Hill:2013hoa} or so-called blind spots in the minimal supersymmetric standard model~\cite{Cheung:2012qy,Huang:2014xua,Crivellin:2015bva}.
More general cases, e.g., retaining a non-vanishing $c_+^M$ as well, are straightforward to consider, but the corresponding limits cannot be represented
in terms of a single-particle cross section anymore. 

In terms of sensitivity to single-particle cross sections, the
coupling to the pion is subleading in chiral EFT with respect to SI,
but dominant over SD scattering. For typical nuclear targets
with $A \sim 100$ nucleons one finds
\beq
\label{scaling}
A^2\gg 4\bigg(\frac{\mpi}{\Lambda_\chi}\bigg)^6\bigg(\frac{\mN}{\mpi}\bigg)^2 A^2 \gg \frac{4}{3}\frac{J+1}{J}\langle {\bf S}_{n/p}\rangle^2,
\eeq
where the middle estimate is for the WIMP--pion coupling, $\Lambda_\chi \sim 500\text{--}600\MeV$ is the chiral EFT breakdown scale, and $\mN$ the nucleon mass. The factor
$(\mpi/\Lambda_\chi)^6$ is due to the subleading $Q^3$ nature of
two-body currents entering quadratically in the cross section---$Q$ is the chiral EFT expansion parameter. For the two xenon isotopes with non-vanishing spin the above
scaling is well reflected by the actual hierarchy of the structure
factors: $1.7\times10^4 \gg 1.1\times10^3 \gg 0.34,0.13$ for
${^{129,131}}$Xe, respectively~\cite{Klos:2013rwa,Hoferichter:2016nvd,Hoferichter:2018acd}.
In this hierarchy, additional contributions from NREFT operators
are further suppressed, because they either vanish at $q=0$
or scale with the very small WIMP velocities $v^2\sim10^{-6}$~\cite{Hoferichter:2016nvd,Hoferichter:2018acd}.
We stress that the scaling~\eqref{scaling} refers to the nuclear responses only, 
so that this hierarchy can always be overcome by a corresponding tuning of the BSM couplings.
In particular, SD~\cite{Aprile:2013doa,Uchida:2014cnn,Aprile:2016swn,Fu:2016ega,Akerib:2017kat} searches probe another complementary slice of the BSM parameter space corresponding to models where SI and WIMP--pion interactions vanish or are strongly suppressed.

\begin{figure}[t] 
\centering
    \includegraphics[width=\columnwidth,clip]{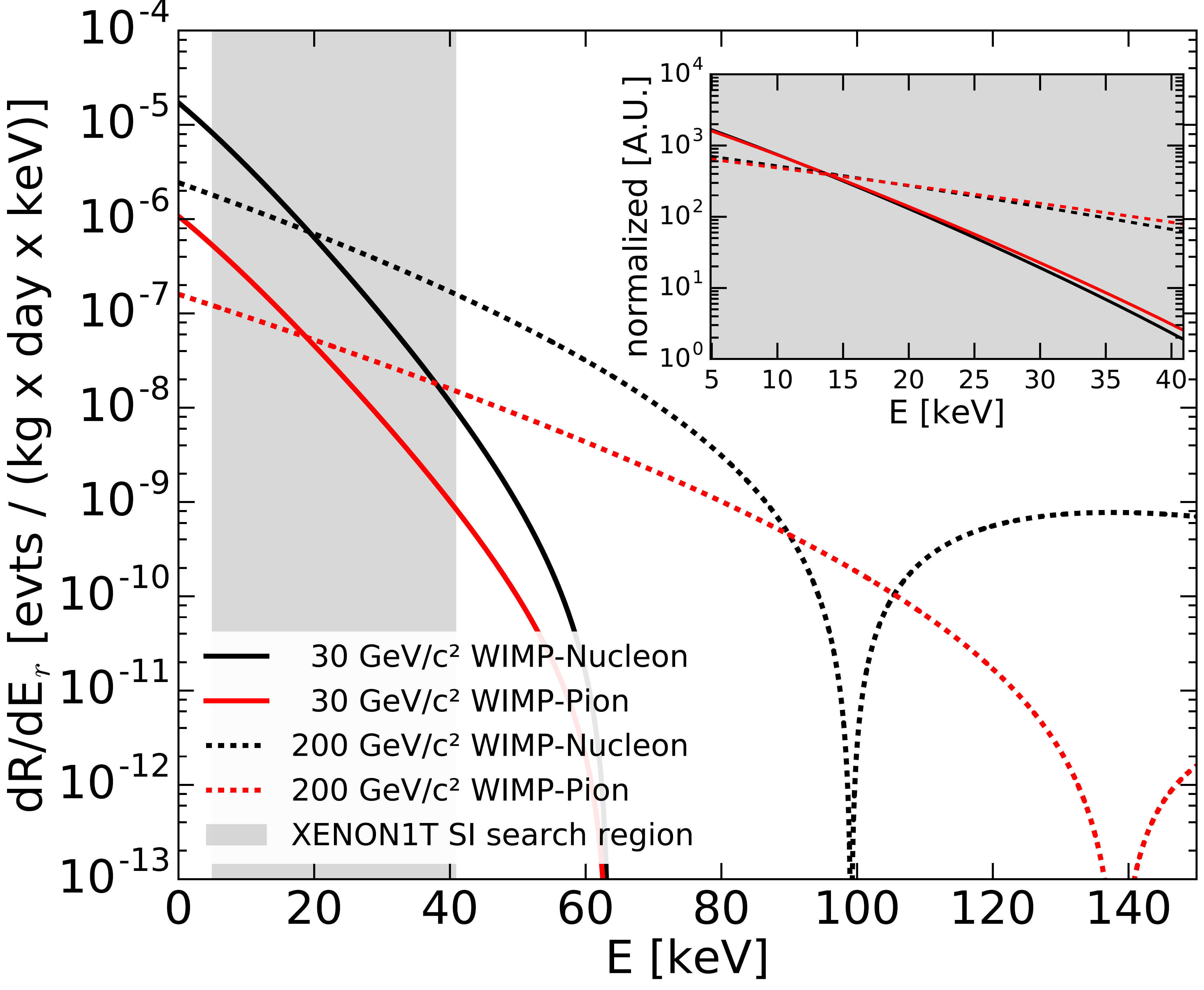}
\caption{Comparison of the differential recoil spectrum for WIMP--nucleon (black) $\F_+^M$ versus WIMP--pion interactions (red) $\F_\pi$. Exemplarily shown are WIMP masses of $30\GeV$ (full line) and $200\GeV$ (dashed line) for the most abundant isotope $^{132}$Xe. The cross section in both cases is set to 10$^{-46}$cm$^2$ for
illustration. The gray band shows the energy range for the XENON1T SI search. The inset compares the spectral shapes in this region.}
\label{fig:dRdE}
\end{figure}

In order to perform the transition from
Eq.~\eqref{eq:SInucleon-coupling} to Eq.~\eqref{eq:pioncoupling} the
signal model has to be adjusted accordingly. For a given WIMP mass,
it is derived from the differential recoil spectrum $\diff
R/\diff E_\text{r}$.  Accounting for the
different kinematic factors in Eq.~\eqref{eq:pioncoupling},
the spectrum for the WIMP--pion coupling can be written as
\beq
\frac{\diff R}{\diff E_\text{r}} = \frac{2\rho_0\sigma_{\chi \pi}^\text{scalar}}{m_{\chi}\mu_\pi^2} \times |\F_\pi(q^2)\big|^2 \times \int_{v_\text{min}(E_\text{r})}^\infty \frac{f(\mathbf{v},t)}{v} \diff^{3}v,
\label{dRdEpion}
\eeq
where $\rho_0$ is the local dark matter density, $f(\mathbf{v},t)$  
its time-dependent velocity distribution truncated at escape velocity, and
$v_\text{min}$ is the minimal WIMP velocity possible for a given recoil energy and detector threshold. 
The main effect of the transition from the SI to the scalar WIMP--pion coupling concerns the form factor, where
$\F_+^M(q^2)$ is replaced by $\F_\pi(q^2)$~\cite{Hoferichter:2016nvd,Hoferichter:2018acd}. Notably, the minimal velocity
remains unchanged as the WIMP is
still scattering off the entire xenon nucleus. A comparison to the standard expression (see Ref.~\cite{Lewin:1995rx}) shows that, as only the form factor influences the shape of the resulting spectrum, both provide a falling
featureless exponential. A comparison of the differential recoil spectra of the WIMP--nucleon and the WIMP--pion scattering is shown in Fig.~\ref{fig:dRdE}. Due to the similarity in shape, the same energy search window can be used for evaluating the WIMP--pion signal model as in
the standard analysis~\cite{Aprile:2018dbl}.
For an attempt to discriminate between SI and WIMP--pion interactions,
see Ref.~\cite{Fieguth:2018vob}.

{\it Experiment}.---To constrain the scalar WIMP--pion coupling, we use data from the XENON1T experiment~\cite{Aprile:2017aty}. 
This data re-analysis is part of the continued use and exploration of the XENON1T 1 t$\times$yr data set. Parallel nuclear recoil (NR) searches are also under way, see Refs.~\cite{Aprile:2013doa,Aprile:2017aas} for XENON100 analyses beyond the SI channel.
We use the same data set and modeling as the SI analysis, except for the signal model, which is replaced by the recoil spectrum in Eq.~\eqref{dRdEpion}. 
The following section gives a brief overview of the XENON1T detector and analysis procedure.

XENON1T is the world's largest dual-phase xenon TPC, shielded by rock overburden at a water equivalent depth of 3600 m at the Laboratori Nazionali del Gran Sasso (LNGS). An active muon veto water tank~\cite{Aprile:2017aty} and an inactive layer of liquid xenon surround the cylindrical TPC. The $2.0$ t target mass of liquid xenon with a gaseous xenon gap at the top is read out by two photomultiplier tube (PMT) arrays, located at the top and bottom of the detector.
Energy deposition within the liquid xenon may produce scintillation photons and ionization electrons. 
Photons are directly registered as the first signal (S1) by the PMTs, while the electrons drift upward in an externally applied field $\mathcal{O}(100\,\text{V/cm})$ to the liquid-gas interface. A strong electric field $\mathcal{O}(10\,\text{kV/cm}$) extracts the electrons into the gas and accelerates them, leading to proportional scintillation in the gaseous phase and thus a secondary light signal (S2).  
The ratio between the two signals (S2/S1) allows one to distinguish statistically between NRs from neutrons and WIMPs, and electronic recoils (ERs) from $\gamma$ and $\beta$ particles. 
The measured S1 and S2 signals are compensated for the spatially inhomogeneous detector response, yielding the corrected analysis variables, cS1 and $\mathrm{cS2_b}$, with the latter measured with the bottom PMT array.  
The time between the prompt S1 signal and the S2 signal measures the depth ($z$-coordinate) of the interaction, while the transversal $(x,y)$ position is reconstructed from the S2 pattern observed by the top PMT array, corrected for a small transverse drift field component. With a three-dimensional position reconstruction of events, the analysis can exclude large background populations at the detector edges by selecting an analysis volume.
Motivated by the similarity of WIMP--pion and SI recoil spectra,
the event selection criteria are the same as in Ref.~\cite{Aprile:2018dbl}. 
The dark matter data is divided into SR0~\cite{Aprile:2017iyp}, with 32.1 days live time, 
and SR1, with 246.7 days. Both the XENON1T SI analysis~\cite{Aprile:2018dbl} and this search use the combined SR0+SR1 data set. 

\begin{figure}[t]
    \centering
    \includegraphics[width=\columnwidth,clip]{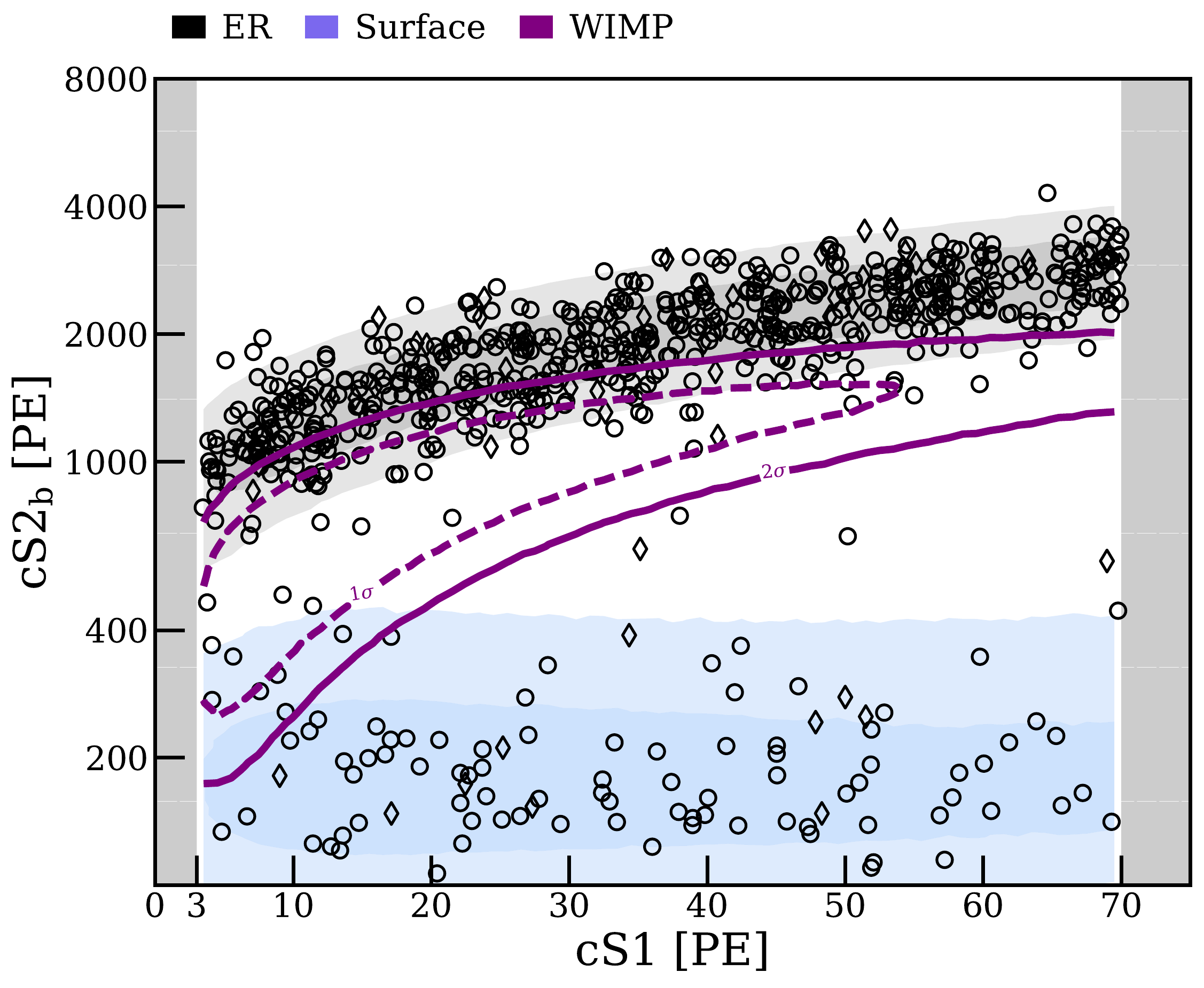}
    \caption{XENON1T SR0+SR1 data (black circles and diamonds, respectively for each period), projected from the three-dimensional analysis space on the primary and secondary scintillation signal, cS1 and $\mathrm{cS2_b}$, in units of photoelectrons (PE). $1\sigma$ and $2\sigma$ containment regions for the WIMP--pion signal model for a $200~\GeV$ WIMP (purple contours), the electronic recoil background (gray bands), and surface background (blue bands) are shown.}
    \label{fig:dataandmodels}
\end{figure}

The signal distribution in cS1 and $\mathrm{cS2_b}$ is derived by convolving the recoil spectrum in Eq.~\eqref{dRdEpion} with the detector NR response, calibrated with a deuterium-deuterium neutron generator~\cite{Lang:2017ymt} and an americium-241-beryllium neutron source. Background distributions for ERs, radiogenic and cosmogenic neutrons, coherent elastic neutrino--nucleus scatters (CE$\nu$NS),
accidental coincidence of
S1 and S2 signals (AC), 
and events originating from the detector surfaces are retained from the SI analysis.  
Figure~\ref{fig:dataandmodels} shows the combined data set, as well as contours for the signal distribution due to a $200\GeV$ WIMP, illustrating 
how a potential signal is separated in the cS1--$\mathrm{cS2_b}$ plane from ER and surface background events. 

Discovery significances and confidence intervals for the WIMP--pion interaction cross section are calculated using the profile likelihood ratio method. The combined likelihood includes extended unbinned likelihood terms for the SR0 and SR1 data sets, using signal and background models in the three-dimensional analysis space (cS1, $\mathrm{cS2_b}$, radius), as well as a core volume with a lower neutron rate~\cite{Aprile:2018dbl}.
The full likelihood also includes additional terms for the ER calibration model fit and ancillary measurements of background rates.
The discovery significance is expressed as the local $p$-value of the observed log-likelihood ratio between the best fit and no-signal models. The null distribution of this parameter is computed for each signal model (WIMP mass) using repeated realizations of the background-only model, since the low signal expectation values preclude the application of asymptotic results.
Confidence intervals, both upper limits and two-sided intervals, are constructed based on a variant of the Feldman--Cousins~\cite{Feldman:1997qc} method using the profile likelihood ratio in the construction of the Neyman band~\cite{Tanabashi:2018oca}. This unified construction avoids undercoverage that can occur when an experiment switches between separate constructions for upper limits and two-sided intervals. The XENON1T experiment places a $3\sigma$ discovery significance threshold for reporting a two-sided interval.

\begin{figure}[t] 
	\centering
    \includegraphics[width=\columnwidth,clip]{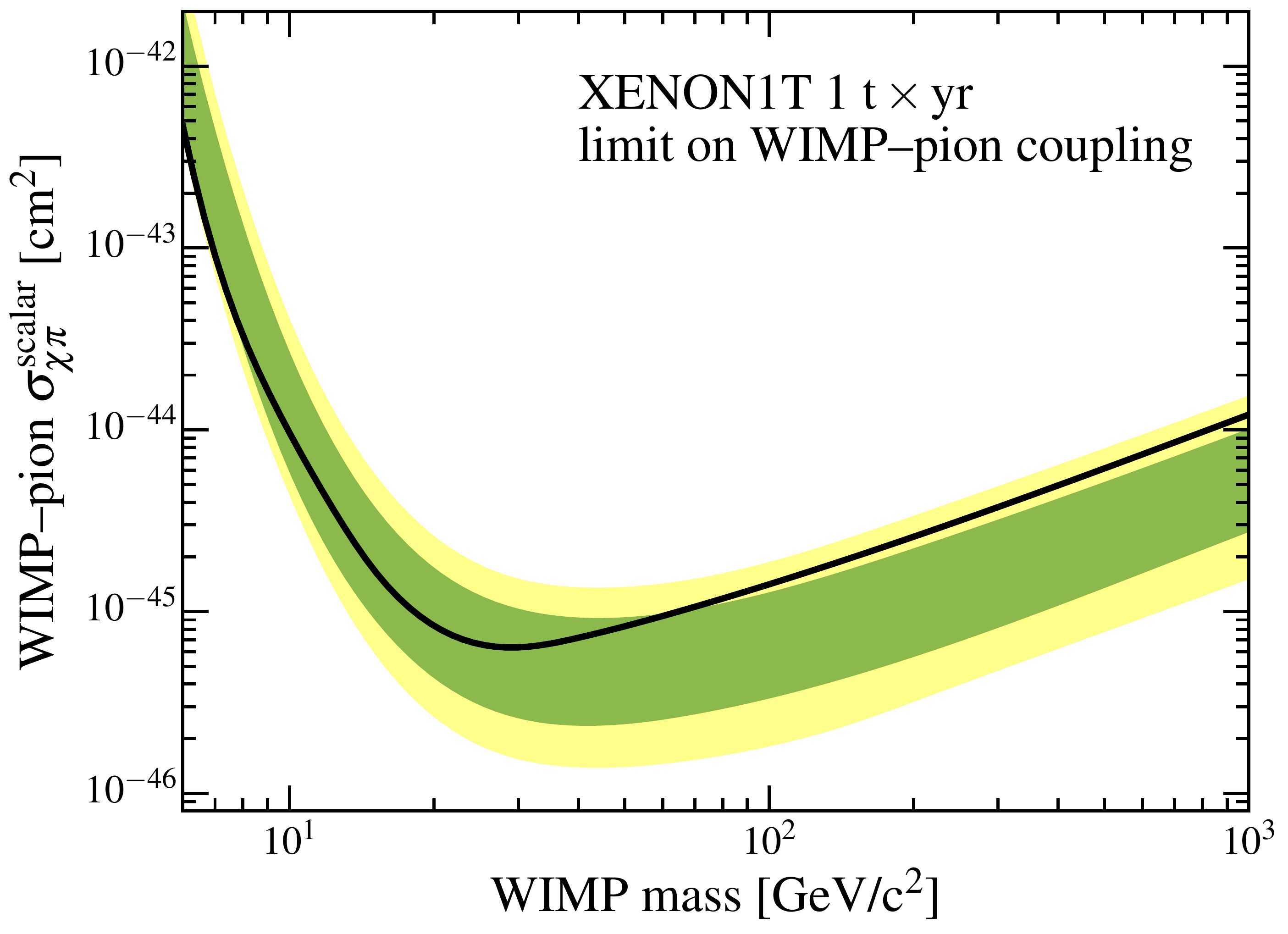}
    \caption{$90\%$ confidence level upper limit of the WIMP--pion coupling as a function of WIMP mass for the 1 t$\times$yr exposure of XENON1T data. Bands show the $1\sigma$ (green) and $2\sigma$ (yellow) quantiles of the expected no-signal distribution of upper limits.
}
	\label{fig:limit}
\end{figure}

\begin{figure}[t] 
	\centering
    \includegraphics[width=\columnwidth,clip]{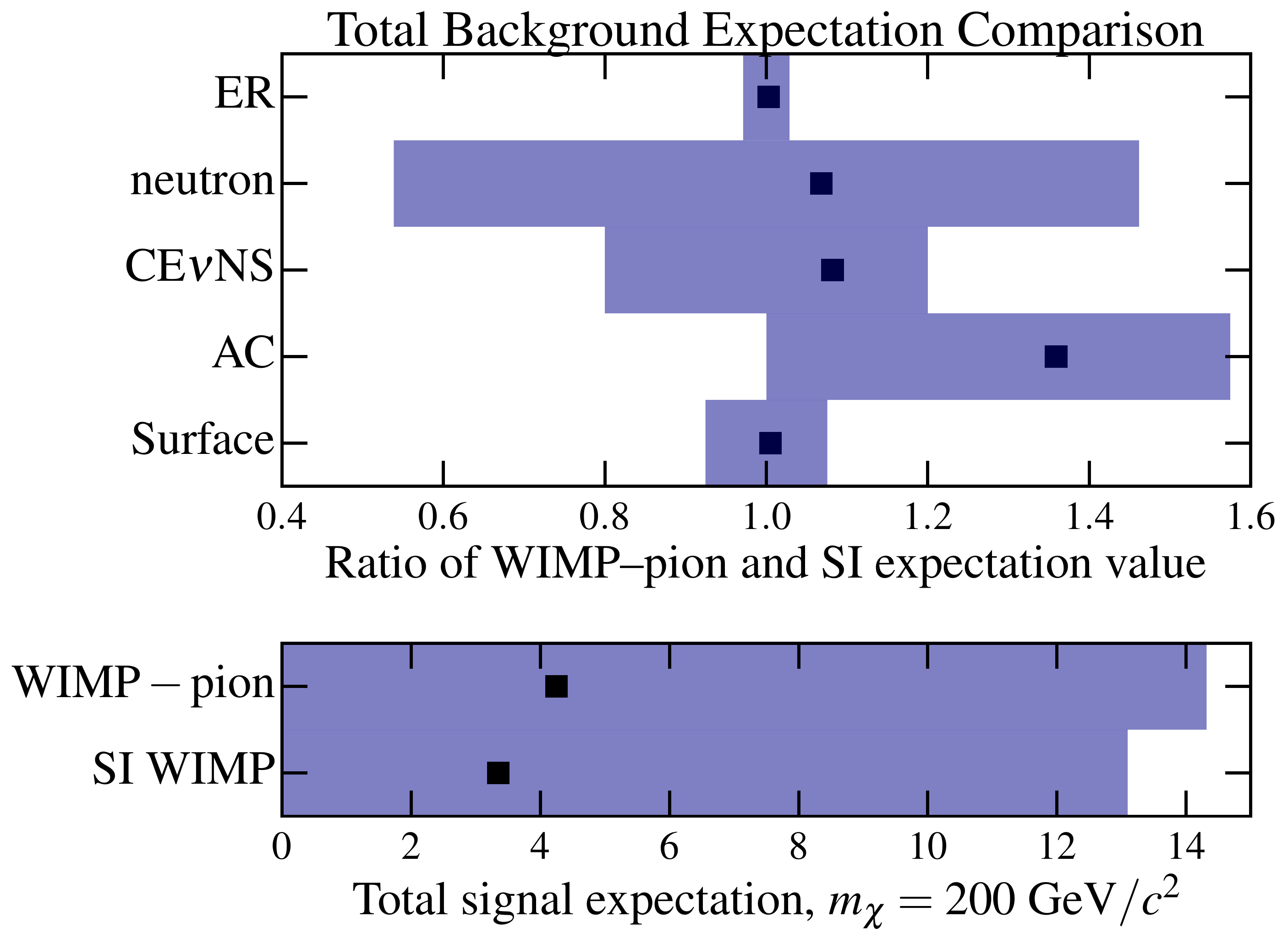}
    \caption{Comparison of the WIMP--pion and standard SI WIMP maximum likelihood estimates (MLEs) of background and signal expectation values. The upper panel shows the ratio between MLE expectation values for this analysis and the SI analysis. Blue bands show the $1\sigma$ confidence bands reported from the SI analysis. The lower panel shows the expected number of signal events for the WIMP--pion and the SI WIMP search~\cite{Aprile:2018dbl} with squares indicating the MLE and the bands the $90\%$ upper limits for the analyses.}
    \label{fig:expectationpulls}
\end{figure}

{\it Result and conclusions}.---No significant signal-like excess is found in our analysis. The lowest local discovery $p$-value is $0.14$, observed for the high mass range above $\sim200\GeV$. The $90\%$ confidence level upper limit on the scalar WIMP--pion cross section, shown in Fig.~\ref{fig:limit}, has a minimum of  $6.4\times10^{-46}~\mathrm{cm}^2$ for a $30\GeV$ WIMP. 
The comparison to the SI analysis is quantified in Fig.~\ref{fig:expectationpulls}, in the upper panel for the ratio of
the background expectations, and in the lower one for the signal 
expectation, both computed for a $200\GeV$ WIMP. No background component shows a significant deviation from the SI fit. 
The upper limits are within $8\%$ in terms of signal expectation value, reflecting the comparable signal recoil energy spectra shown in Fig.~\ref{fig:dRdE}. The difference in upper limit cross sections is therefore driven primarily by the different expectation values for the two interactions at the same cross section.

Summarizing, we have presented limits on the scalar
WIMP--pion interaction, where the WIMP scatters off
virtual pions in a nucleus via an underlying scalar
WIMP--quark operator. The corresponding nuclear response
for this interaction is coherently enhanced,
similarly to SI scattering, leading to the hierarchy given in Eq.~\eqref{scaling}.
In analogy to standard SI and SD limits,
the result can be represented in terms of a single-particle
cross section.  
We have performed the first search for this interaction with 1 t$\times$yr of XENON1T
data, using the XENON1T detector response, background models, and likelihood.
We find no excess and set an upper limit on the scalar WIMP--pion cross section with a minimum at $6.4\times10^{-46}~\mathrm{cm}^2$ for a $30\GeV$ WIMP (at 90\% confidence level). 
Our analysis quantifies for the first time the effect of
coherent two-body currents in direct-detection searches
for dark matter, paving the way for future comprehensive studies of WIMP--nucleus interactions beyond SI
and SD scattering.

\begin{acknowledgments}
    The XENON collaboration gratefully acknowledges support from the National Science Foundation, Swiss National Science Foundation, German Ministry for Education and Research, Max Planck Gesellschaft, Deutsche Forschungsgemeinschaft, Netherlands Organisation for Scientific Research (NWO), Netherlands eScience Center (NLeSC) with the support of the SURF Cooperative, Weizmann Institute of Science, Israeli Centers Of Research Excellence (I-CORE), Pazy-Vatat, Initial Training Network Invisibles (Marie Curie Actions, PITNGA-2011-289442), Fundacao para a Ciencia e a Tecnologia, Region des Pays de la Loire, Knut and Alice Wallenberg Foundation, Kavli Foundation, Abeloe  Fellowship, and Istituto Nazionale di Fisica Nucleare. Data processing is performed using infrastructures from the Open Science Grid and European Grid Initiative. We are grateful to Laboratori Nazionali del Gran Sasso for hosting and supporting the XENON project.

    The work of M. H., P. K., J. M., and A.~S.~was supported by the US DOE (Grant No.\ DE-FG02-00ER41132), the ERC (Grant No.\ 307986 STRONGINT), the Deutsche Forschungsgemeinschaft through SFB 1245 (Projektnummer 279384907), the Max-Planck Society, the Japanese Society for the Promotion of Science KAKENHI (Grant 18K03639), MEXT  as "Priority Issue on Post-K computer" (Elucidation of the fundamental laws and evolution of the universe), JICFuS  and the CNS-RIKEN joint project for large-scale nuclear structure calculations.
\end{acknowledgments}

\end{document}